\documentclass{ectjeditorial}
\usepackage{amssymb,amsmath,dsfont}
\usepackage{tikz}
\usetikzlibrary {positioning}

\renewcommand{\Pr}{\mathbb{P}}

\newcommand{\ad}{\overset{{\mathrm{a}}}{\sim}}
\newcommand{\Ep}{{\mathbb{E}}}
\newcommand{\En}{\mathbb{E}_{n}}
\newcommand{\Vn}{\mathbb{V}_{n}}
\newcommand{\mP}{\mathcal{P}}
\newcommand{\mQ}{\mathcal{Q}}
\newcommand{\indep}{\perp\!\!\!\!\perp}

\newcommand{\cstoZ}{K_1}
\newcommand{\cstoDS}{K_2}

\DeclareMathOperator{\Var}{Var}

\newtheorem{algorithm}{Algorithm}
\newtheorem{remark}{Remark}

\firstheader{Editorial prepared for {\em The Econometrics Journal 28}(1), https://doi.org/10.1093/ectj/utaf006}
\received{January 2025}
\volume{28}
\title[Editorial]{Philip G. Wright, directed acyclic graphs, and instrumental variables}
\author[J. H. Abbring, V. Chernozhukov, and I. Fern\'{a}ndez-Val]{}

\begin{document}
\begin{center}
    \textsc{EDITORIAL}\\ \ \\
\end{center}

\noindent This Special Issue celebrates the seminal contributions of Philip G. Wright to causal inference in economics and puts them in a modern perspective. At its core is a reproduction of Appendix B in Wright's 1928 book {\em The Tariff on Animal and Vegetable Oils}.\footnote{Wright, P. G. (1928). {\em The Tariff on Animal and Vegetable Oils}. New York: The Macmillan Company.} This book deals with demand for and supply of oils and butter. In its Appendix B,  Wright made several fundamental contributions to causal inference. He introduced a structural equation model of demand and supply, established the identification of demand and supply elasticities via the method of moments and directed acyclical graphs (DAGs), developed empirical methods for estimating demand elasticities using weather conditions as instruments, and proposed methods for counterfactual analysis of the welfare effects of imposing tariffs and taxes. Moreover, he took all of these methods to data. 

Wright's (1928) Appendix B was far ahead of, and much more profound than, any contemporary theoretical and empirical developments on causal inference in statistics or econometrics. In the years leading up to its publication, economists had developed an appreciation for the identification problem in economics, with a focus on identifying demand and supply curves from price and quantity data.\footnote{For historical accounts of these early studies of the identification problem, see Chapter 6 in Morgan, M.S. (1990). {\em The History of Econometric Ideas}. Cambridge: Cambridge University Press; and the introduction to Part III, on pages 17--33, in Hendry, D. F. and M. S. Morgan (1995). {\em The Foundations of Econometric Analysis}. Cambridge: Cambridge University Press.}  Wright himself was one early contributor to this literature, with his 1915 review of Henry L. Moore's book {\em Economic Cycles}.\footnote{Moore, H. L. (1914). {\em Economic Cycles; Their Law and Cause}. New York: The Macmillan Company.} In this book, Moore provided an extensive analysis of one important source of economic fluctuations, rainfall, and subsequently analysed the statistical relations between percentage changes in observed prices and quantities of various goods over time.  Moore argued that these `statistical laws of demand' described `average changes that society is actually undergoing' (page 77) and were therefore more useful for prediction than Marshallian demand curves, with their dependence on abstract {\em ceteris paribus} conditions. He moreover emphasized that, for some goods, his statistical laws of demand substantially differed from typical Marshallian demand curves (page 126):
\begin{quote}
Unlike the law of demand for the crops, the law of demand for a representative producers’ good is such that as the supply increases the price rises, and as the supply decreases the price falls. 
\end{quote}
In his 1915 review of Moore's book, Wright noted that rainfall is likely to shift the supply of crops but, via its effect on business conditions, the demand for a producers' good like pig iron.\footnote{Wright, P. G. (1915). Moore’s economic cycles. {\em The Quarterly Journal of Economics 29}(2), 631--41.} He concluded that Moore's downward sloping statistical laws of demand for crops may well coincide with their Marshallian demand curves, but that Moore's upward sloping law of statistical demand for pig iron may be closer to a {\em ceteris paribus} supply curve. Wright also noted that {\em ceteris paribus} demand and supply curves may be more useful for {\em counterfactual} predictions than Moore's statistical laws of demand (page 639): 
\begin{quote}
Suppose, for example, we were to accept as universal the inductive law of producers’ goods given on page 114. “The price rises with an increase of the product and falls with its decrease”; and suppose, furthermore, that manufacturers of pig iron on the strength of this “universal law” should deliberately double, treble, or quadruple their output in the confident expectation that prices would rise proportionately: I fear that thereafter Professor Moore would not stand high as a prophet among producers of pig iron. 
\end{quote}
Both Wright's (1915) and contemporaneous analyses of the identification problem primarily relied on diagrammatic and verbal arguments. Wright's (1928) Appendix B stands out by developing formal econometric methods for a structural model of demand and supply. In particular, it is generally credited with introducing instrumental variables (IV) estimation.\footnote{As Philip Wright's Appendix B is quite different from the rest of his book and relies in part on the path analysis approach developed by his son, Sewall Wright, some have argued that Sewall was in fact the author of Appendix B. However, stylometric analysis points to Philip Wright as the author. See Stock J. H. and F. Trebbi (2003). Who invented instrumental variable regression? {\em Journal of Economic Perspectives 17}(3), 177--94.}

Sections \ref{s:WrightsSEM}--\ref{s:contemporary} of this editorial provide a modern perspective on Wright's work in a lecture note format that can be useful for teaching. Section \ref{s:outline} comments on our reproduction of Wright's Appendix B and introduces the three original articles that follow it.

\section{Wright's System of Demand and Supply Equations}
%===================================================================================
\label{s:WrightsSEM}

We begin with a stochastic model of demand and supply for a given commodity. Consider the system 
\begin{equation} \label{eqn:Wright} 
  \begin{aligned} 
   && D(p) &:= \alpha_1 p + U^d\text{ and}&&\\
   && S(p) &:= \beta_1 p + U^s, &&
  \end{aligned}
\end{equation}
where $D(p)$ denotes the (log) quantity demanded at the (log) price level $p$,   $S(p)$ denotes the (log) quantity supplied at the (log) price level $p$, and the random variables $U^d$ and $U^s$ are stochastic factors that shift demand and supply and are not affected by $p$.  Thus, the maps $p \mapsto  D(p)$ and $p \mapsto  {S(p)}$ are random aggregate demand and supply curves and $p \mapsto \alpha_1 p$ and $p \mapsto \beta_1 p$ describe the deterministic parts of these curves. Because all variables are in logarithms, the parameters $\alpha_1$ and $\beta_1$ correspond to the demand and supply price elasticities. 

Suppose further  that the stochastic factors can be decomposed into observable and unobservable parts:
\begin{equation} \label{eqn:U}
  \begin{aligned}     
 &U^d:= \alpha_2' Z^d + \alpha_3'W + \epsilon^d, \quad && \epsilon^d \perp Z^d, Z^s, W;\text{ and}\\
 &U^s := \beta_2' Z^s + \beta_3'W + \epsilon^s, \quad      && \epsilon^s \perp Z^d, Z^s, W;
  \end{aligned}
\end{equation}
where the components of the random vector $W$ are common shifters of demand and supply (that include a constant), components of $Z^d$ shift only demand, and components of $Z^s$ shift only  supply. The residual shocks $\epsilon^d$ and $\epsilon^s$ capture the effects of all the unobservable shifters of the demand and supply curves, which are orthogonal to all the observable shifters by construction.  

To clarify the role of the observable factors in explaining the fluctuations of the demand and supply curves, we can think of \eqref{eqn:U} as a pair of linear regression equations. The choice of which variables to include in $W$, $Z^d$, and $Z^s$ is similar to specification analysis in regression.  It is critical in Wright's empirical strategy to find strong demand and supply shifters, which requires careful contextual reasoning and good data collection.  For example, the supply of crops is affected by weather conditions, so we need to collect weather data at appropriate locations.
 
We observe the Walrasian equilibrium price $P$ of the demand and supply system \eqref{eqn:Wright}, the price $p$ that sets the quantity $D(p)$ demanded equal to the quantity $S(p)$ supplied, and the corresponding equilibrium quantity $Y:=D(P)=S(P)$. In modern causal language, $D(p)$ and $S(p)$ are `potential' demand and supply outcomes, the quantities that would be demanded and supplied if the price would be fixed at $p$, and the observed outcome $Y$ is obtained by evaluating $D(p)$ or $S(p)$ at the observed price $p=P$. Once equilibrium outcomes $(P,Y)$ are realized, demand $D(p)$ and supply $S(p)$ at prices $p\neq P$ are `counterfactual'.\footnote{In 1923, Neyman introduced potential outcomes in the context of agricultural experiments. Neyman's work was published in Polish and not internalized until Rubin singled out the central role of potential outcomes in formulating causal inference problems. See Neyman, J. (1990). On the application of probability theory to agricultural experiments. Essay on principles. Section 9 (D. M. Dabrowska and T. P. Speed, Trans.). {\em Statistical Science 5}(4), 465--72. (Original work published 1923); and Rubin, D. B. (1974). Estimating causal effects of treatments in randomized and nonrandomized studies. {\em Journal of Educational Psychology 66}(5), 688–70. 

In contrast, in the early econometric literature, Wright, but also Tinbergen and Haavelmo, explicitly emphasized potential and possibly counterfactual outcomes as naturally arising from structural economic models. See {\em e.g.} Tinbergen, J. (1930). Bestimmung und Deutung von Angebotskurven: ein Beispiel. {\em Zeitschrift f\"{u}r  National\"{o}konomie 1}, 669--79; and Haavelmo, T. (1943). The statistical implications of a system of simultaneous equations. {\em Econometrica 11}(1), 1--12. 

Finally, note that Wright did not explicitly use the Rubin-style notation $D(p)$ and $S(p)$, but did plot the demand and supply curves $p\mapsto D(p)$ and $p\mapsto S(d)$.}

Substituting the equilibrium price $P$ and quantity $D(P)=S(P)=Y$ in the original equations yields an econometric simultaneous equations model:
\begin{equation}\label{eqn:SEM}%\tag{Wright's SEM}
\begin{aligned}
&Y - \alpha_1 P = \alpha_2' Z^d + \alpha_3'W + \epsilon^d, \quad &&\epsilon^d \perp \left(Z^d, Z^s, W\right);\text{ and} \\
&Y- \beta_1 P  = \beta_2' Z^s + \beta_3'W + \epsilon^s, \quad &&\epsilon^s \perp \left(Z^d, Z^s, W\right).
\end{aligned}
\end{equation}
Here we collect the endogenous variables--- those determined by the model --- on the left-hand side and the exogenous variables --- those determined outside the model--- on the right side.  However, we can put variables on any side we want, because the equations in \eqref{eqn:SEM} are structural, with orthogonality restrictions on the stochastic shocks, and
{\em not} regression equations.  
\begin{figure}[t]   
    \vspace*{10pt}
    \begin{center}
    \begin{tikzpicture}[scale=0.8]
	\draw[line width=1pt]  (0,0) -- (0,10.5);        
	\draw[line width=1pt] (0,0) -- (11.5,0)  node[anchor= north east] {$p$};   
        \draw[line width=1pt,color=blue] (1,8) --  (8,2) node[anchor= north west, black] {$D(p)(\omega_1)$};
        \draw[line width=1pt,color=blue] (2,9) --  (9,3) node[anchor= north west, black] {$D(p)(\omega_2)$};
        \draw[line width=1pt,color=blue] (1,6) --  (7,1) node[anchor= north west, black] {$D(p)(\omega_3)$};
       
        \draw[line width=1pt,color=red] (2,2) -- (8,8) node[anchor=south west, black] {$S(p)(\omega_1)$};
        \draw[line width=1pt,color=red] (4.5,1.5) -- (9.5,6.5)node[anchor=south west, black] {$S(p)(\omega_2)$};
        \draw[line width=1pt,color=red] (1.5,4.5) -- (6.5,9.5)node[anchor=south west, black] {$S(p)(\omega_3)$};

	\draw[fill] (4.7692,4.7692) circle [radius=.1];
	\draw[fill] (2.0909,5.0909) circle [radius=.1];
	\draw[fill] (7.3846,4.3846) circle [radius=.1];
    \end{tikzpicture}
    \end{center}
 \footnotesize \textbf{Note:} This figure plots demand and supply curve realizations for three different states of the world; $\omega_1$, $\omega_2$, and $\omega_3$; which can either correspond to different periods or different markets. The equilibrium values $(P,Y)$ are the points of intersection of the pairs of curves $p \mapsto D(p)(\omega_t)$ and $p \mapsto S(p)(\omega_t)$; $t=1,2,3$.
 
 \caption{Three different realizations of random demand and supply curves.\label{fig:supply-demand}}
 \end{figure}
 
  \begin{figure}[ht]
    \vspace*{10pt}
    \begin{center}
    \begin{tikzpicture}[scale=0.8]
	\draw[line width=1pt]  (0,0) -- (0,10.5);        
	\draw[line width=1pt] (0,0) -- (11.5,0)  node[anchor= north east] {$p$};  
	\draw[line width=1pt,color=blue] (1,8) --  (9,1) node[anchor= north west, black] {$D(p)$};
               
        \draw[line width=1pt,color=red] (2,2) -- (8,8) node[anchor=south west, black] {$S(p)(\omega_1)$};
        \draw[line width=1pt,color=red] (4.5,1.5) -- (9.5,6.5)node[anchor=south west, black] {$S(p)(\omega_2)$};
        \draw[line width=1pt,color=red] (1.5,4.5) -- (6.5,9.5)node[anchor=south west, black] {$S(p)(\omega_3)$};

	\draw[fill] (4.7333,4.7333) circle [radius=.1];
	\draw[fill] (6.3333,3.3333) circle [radius=.1];
	\draw[fill] (3.1333,6.1333) circle [radius=.1];
    \end{tikzpicture}
 \end{center}   
     \footnotesize \textbf{Note:} This figure plots demand and supply curve realizations for three different states of the world; $\omega_1$, $\omega_2$, and $\omega_3$; where $D(p)(\omega_1)=D(p)(\omega_2)=D(p)(\omega_3)=:D(p)$. The equilibrium values $(P,Y)$ are the points of intersection of the curve $p \mapsto D(p)$ with the curves $p \mapsto S(p)(\omega_t)$; $t=1,2,3$.
    
    \caption{Three realizations of the demand and supply curves where the supply curves shift, but the demand curve remains unchanged.\label{fig:supply-demand2}}
\end{figure}
 
Wright noted that there is no hope to identify the structural elasticity parameters $\alpha_1$ and $\beta_1$ from the linear projection coefficients of $Y$ on $P$ or $P$ on $Y$.  With the aid of plots like Figure \ref{fig:supply-demand}, he argued that equilibrium prices and quantities would not necessarily trace either the demand or the supply curve.  However, as illustrated in Figure \ref{fig:supply-demand2}, we can identify the demand curve if we can isolate cases where the supply curve shifts while the demand curve remains unchanged.\footnote{Interestingly, this is a heuristic identification strategy that Giacomini, Kitagawa, and Read recently formalized as using `narrative' restrictions in identification.  In this approach, an analyst could also use known directional movements in demand and supply curves relative to the base period to bound elasticities. The case of demand staying constant, enabling point identification, is a particular, extreme case. See Giacomini, R., T. Kitagawa, and M. Read (2021). Identification and inference under narrative restrictions. arXiv:2102.06456 {\em [econ.EM]}.}  Well chosen instrumental variables can act as such supply shifters, with the added difficulty that the demand curve is also moving at the same time, but in directions that are uncorrelated with the movements of the supply curve.  When supply shifters are available, we conclude that the data contain some `quasi-experimental' variation  that can be used to trace out the typical shape of the demand curve and thus identify the demand elasticity $\alpha_1$.  Similarly, demand shifters can be used to identify the supply elasticity $\beta_1$.  Below, we show how to identify and estimate these and other parameters. 

\begin{figure}[t]
\vspace*{10pt}
\begin{center}
\begin {tikzpicture}[-latex, auto, node distance =1.5cm and 3cm, on grid, thick, 
  observed/.style ={circle, top color=white, bottom color = blue, draw, black, text=black , minimum width =1.75 cm},
  unobserved/.style ={circle, top color=white, bottom color = yellow, draw, black, text=black , minimum width =1.75 cm}, 
  eqb/.style ={rectangle, top color=white, bottom color = magenta, draw, black, text=black , minimum width =1.75 cm}, 
empty/.style ={circle, top color=white, bottom color = white, draw,  white, text= white, minimum width =.5 cm}]

\node[unobserved]  (Yd) at (2,2)  {$D(\bullet)$};
\node[unobserved]   (Ys) at (2,-2) {$S(\bullet)$};
\node[observed]   (Zs) at (-2,-2) {$ Z^s$};
\node[observed]   (Zd) at (-2,2) {$ Z^d$};
\node[observed]    (P) at (6,2) {$P$};
\node[observed]    (Y) at (6,-2) {$Y$};
\path (Ys) edge (Y);
\path (Yd) edge (Y);
\path (Ys) edge (P);
\path (Yd) edge (P);
\path (Zd) edge (Yd);
\path (Zs) edge (Ys);
\node[unobserved]  (L2) at (0,0)  {$\cstoDS$};
\node[unobserved]  (L1) at (-4,0)  {$\cstoZ$};
\path (L1) edge (Zd);
\path (L1) edge (Zs);
\path (L2) edge (Yd);
\path (L2) edge (Ys);
\end{tikzpicture} 
\end{center}
{\footnotesize \textbf{Note:} Nodes represent random elements (variables, vectors, or functions). Observed nodes are blue (shaded in grayscale) and latent nodes are yellow (clear).  Directed edges represent causal effects of parent nodes on child nodes. The nodes are distributed as a Markov network: Each node's distribution depends only on its immediate parents and not on any other nondescendant nodes.  The absence of links between nodes implies their independence (lack of correlation in linear models) conditional on their parents (a local Markov condition).  $Z^d$ is a vector of demand shifters, $ Z^s$ a vector of supply shifters,  $D(\cdot)$ a stochastic demand function, and $S(\cdot)$ a stochastic supply function. The demand $D(\cdot)$ and supply $S(\cdot)$ determine the equilibrium price $P$ and quantity $Y$. The latent $\cstoZ$ is a common determinant of $Z^d$ and $Z^s$ and makes explicit that these may be dependent (correlated). Similarly, the latent $\cstoDS$ is a common determinant of $D(\cdot)$ and $S(\cdot)$. As there is no edge between $\cstoZ$ and $\cstoDS$ and neither has parents, they are independent (uncorrelated). As there is no edge from $ Z^s$ to $D(\cdot)$,  $Z^s$ is independent of (uncorrelated with) $D(\cdot)$, after controlling for $ Z^d$.  Likewise, as there is no edge from $ Z^d$ to  $S(\cdot)$, $Z^d$ is independent of (uncorrelated with) $S(\cdot)$, after controlling for  $ Z^s$. This DAG is a version of Wright's (1928) Figure 10 that shows the random demand and supply curves $D(\cdot)$ and $S(\cdot)$ instead of Wright's overall demand and supply shocks $U^d$ and $U^s$ and that, unlike Wright, includes the latent nodes $\cstoZ$ and $\cstoDS$ to account for dependence (correlation) of the demand and supply shifters $Z^d$ and $Z^s$ and of the other determinants of demand and supply. We don't show $W$ to keep the graph simple.  We could place it as a node in the graph with edges $W \to S(\cdot)$, $W \to D(\cdot)$, and $\cstoZ\to W$. We have reproduced Wright's original figure in Figure \ref{fig:DAG2}.}

\caption{A (modernised) causal diagram of Wright's model.\label{fig:DAG}} 
\end{figure}
%====

Philip Wright was also a pioneer in applying causal path diagrams to analyze identification. He employed causal path diagrams (now called causal DAGs), introduced earlier by his son, Sewall Wright. In Figure \ref{fig:DAG} we present a slightly more general, nonparametric version of Wright's (1928, Figure 10) causal DAG.   One impressive feature of this figure is that it is one of the earliest applications of DAGs to causal inference.\footnote{Other examples are due to his son Sewall; see {\em e.g.} Wright, S. (1923). The theory of path coefficients: a reply to Niles's criticism. {\em Genetics 8}(3), 239--55.} Another impressive feature is that this figure gives an {\em acyclical} representation of the simultaneous equations model, which is distinct from the standard cyclical representations in {\em e.g.} Pearl's work.\footnote{\label{fn:cyclical}See {\em e.g.} Section 7.2.1 and Figure 7.4 in Pearl, J. (2009b). {\em Causality: Models, Reasoning, and Inference}, 2nd edn. Cambridge: Cambridge University Press. Specifically, the cyclical causal diagram takes the form: $Z^d \to Y$, $Z^d \to P$,  $Z^s \to Y$, $Z^s \to P$, and  $Y \leftrightharpoons P$; see Richardson, T. S. and P. Spirtes (1996). Automated discovery of linear feedback models. Report CMU-PHIL-75, Philosophy Department, Carnegie Mellon University; and White, H. and K. Chalak (2009). Settable systems: an extension of Pearl’s causal model with optimization, equilibrium, and learning. {\em Journal of Machine Learning Research 10}(8).} This uses that cycles are not needed once we allow for latent variables in the diagram, potential demand and supply. Given the potential demand and supply nodes, the Walrasian auctioneer computes the  price that clears the market and this equilibrium price and the resulting quantity traded are observed.\footnote{\label{fn:finecycles}In principle, there is nothing wrong with cycles either, as long as we properly formalize them; see Richardson and Spirtes (1996) and White and Chalak (2009) in Footnote \ref{fn:cyclical}. Furthermore, a simultaneous equations model may correspond to the limit of an acyclical dynamic system, as in Fisher, F. M. (1970). A correspondence principle for simultaneous equation models. {\em Econometrica 38}(1), 73–92.  This `correspondence principle' does not directly apply to Wright's simultaneous equations model, which we may however interpret as the limit outcome of an acyclical Walrasian `tat\^{o}nnement' price adjustment process towards equilibrium; for discussion, see Richardson, T. S. and J. M. Robins (2014). ACE bounds; SEMs with equilibrium conditions. {\em Statistical Science 29}(3), 363–66.}  

The structure of the DAG in Figure \ref{fig:DAG} also encodes the orthogonality assumptions in Wright's simultaneous equations model \eqref{eqn:SEM}.  For example, the demand is uncorrelated with the supply shifters after controlling linearly for the demand and the common shifters:
$$
(D(p))_{p \in \Bbb{R}} \perp Z^s \mid Z^d, W.
$$
This conclusion follows from observing that working with $(D(p))_{p \in \Bbb{R}}$  is equivalent to working with the demand shock $U^d$. If we strengthen the absence of links in the DAG to signify independence of nodes conditional on parental nodes, then the DAG encodes stronger restrictions, such as that, after conditioning on $Z^d$ and $W$, $Z^s$ is independent of the demand:
$$
(D(p))_{p \in \Bbb{R}} \indep Z^s \mid Z^d, W.
$$
This is the conventional conditional exogeneity or ignorability condition widely employed in modern econometrics and statistics.  Thus, Wright's work is an important precursor to modern analysis in statistics, machine learning, and econometrics, which relies on DAGs to encode contextual knowledge and deduce conditional exogeneity conditions.

\section{Wright's Identification Arguments} 
%===================================================================================

Here we simplify the arguments to get back to the original formulation of Wright.  We assume that there is no $W$ (or that it has been partialed out, as we do in the next section), and that $Z^d$ and $Z^s$ are scalar random variables.  The crux of Wright's method lies in the realization that the following moment conditions hold as the result of the assumed orthogonality conditions in Wright's simultaneous equations model \eqref{eqn:SEM}:
\begin{equation}
\label{key}
\begin{aligned}
&\Ep ( Y - \alpha_1  P - \alpha_2  Z^d) ({ Z^s}, { Z^d})'=0\text{ and}\\
&\Ep ( Y - \beta_1 P - \beta_2 Z^s) ({ Z^s}, {Z^d})' = 0.
 \end{aligned}
\end{equation}
We further assume that, as in Wright's original analysis, $Z^d$ and $Z^s$ are uncorrelated and centered. In this case, \eqref{key} implies
\begin{equation}\label{key-2}
\Ep ( Y - \alpha_1  P ) { Z^s} = 0\quad\text{and}\quad\Ep ( Y - \beta_1 P )  {Z^d} = 0.
 \end{equation}
Solving this system of equations, we recover the IV formulas 
$$
\alpha_1 =  \frac{\Ep ( Y Z^s)}{\Ep ( P   Z^s) }\quad\text{and}\quad
\beta_1 =  \frac{\Ep ( Y Z^d)}{\Ep ( P   Z^d)},
$$
which identify the demand and supply elasticities $\alpha_1$ and $\beta_1$, provided that the denominators are not equal to zero; that is, the supply and demand shifters have a nontrivial effect on the equilibrium price.  This is the instrument relevance condition in econometrics.  The above gives a `method of moments' version of Wright's main argument.

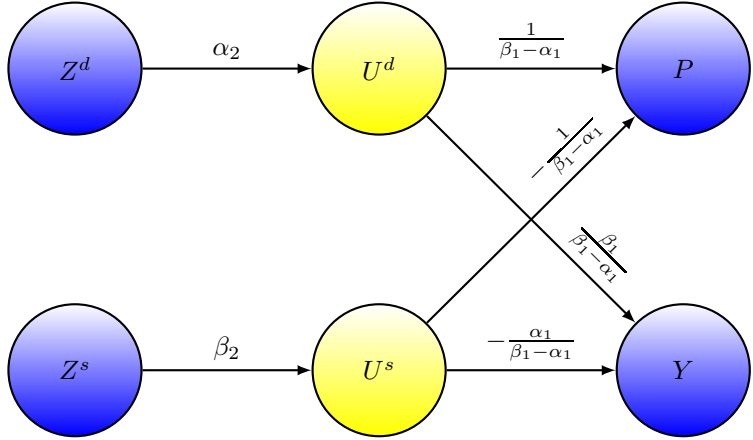
\begin{figure}[t]
\vspace*{10pt}
\begin{center}
\begin {tikzpicture}[-latex, auto, node distance =1.5cm and 3cm, on grid, thick, 
  observed/.style ={circle, top color=white, bottom color = blue, draw, black, text=black , minimum width =1.75 cm},
  unobserved/.style ={circle, top color=white, bottom color = yellow, draw, black, text=black , minimum width =1.75 cm}, 
  eqb/.style ={rectangle, top color=white, bottom color = magenta, draw, black, text=black , minimum width =1.75 cm}, 
empty/.style ={circle, top color=white, bottom color = white, draw,  white, text= white, minimum width =.5 cm}]

\node[unobserved]  (Yd) at (2,2)  {$U^d$};
\node[unobserved]  (Ys) at (2,-2) {$U^s$};
\node[observed]   (Zs) at (-2,-2) {$Z^s$};
\node[observed]   (Zd) at (-2,2) {$Z^d$};
\node[observed]    (P) at (6,2) {$P$};
\node[observed]    (Y) at (6,-2) {$Y$};
\path (Ys) edge node [above]{ $-\frac{\alpha_1}{\beta_1 - \alpha_1}$}  (Y);
\path (Yd) edge  node [pos=0.75,above,rotate=-45] {$\frac{\beta_1}{\beta_1 - \alpha_1}$} (Y);
\path (Ys) edge node [pos=0.75,above,rotate=45] {$-\frac{1}{\beta_1 - \alpha_1}$} (P);
\path (Yd) edge node [above]{ $\frac{1}{\beta_1 - \alpha_1}$}  (P);
\path (Zd) edge node [above]{$\alpha_2$} (Yd);
\path (Zs) edge node [above]{$\beta_2$} (Ys);
\end{tikzpicture} 
\end{center}
{\footnotesize \textbf{Note:} Nodes represent random variables. Observed nodes are blue (shaded in grayscale) and latent nodes are yellow (clear).  Directed edges represent linear causal relations. The absence of links between nodes signifies their lack of correlation, given their parents.  $Z^d$ is a demand shifter, $Z^s$ a supply shifter, $U^d$ a stochastic component of demand, and $U^s$ a stochastic component of supply. The demand and supply components $U^d$ and $U^s$ determine the equilibrium price $P$ and quantity $Y$.  The absence of a common determinant $\cstoZ$ of $Z^d$ and $Z^s$ (as in Figure \ref{fig:DAG}) and the lack of an edge between them together reflect the simplifying assumption that they are uncorrelated. The absence of a common determinant $\cstoDS$ of $U^d$ and $U^s$ (as in Figure \ref{fig:DAG}) and the lack of an edge between them together reflect the simplifying assumption that they are uncorrelated (given $Z^d$ and $Z^s$; conditioning has no bite here and in the next two statements because  $Z^d$ and $Z^s$ are uncorrelated and the model is linear). As there is no edge from $Z^s$ to  $U^d$, $Z^s$ is uncorrelated with $U^d$ (given $Z^d$). Similarly, as there is no edge from  $Z^d$ to $U^s$, $Z^d$ is uncorrelated with $U^s$ (given $Z^s$).  }

\caption{A version of Wright's (1928) Figure 10.\label{fig:DAG2}}
\end{figure}

We next get to Wright's graphical derivations of the IV formula, again using the simplifying assumption that $Z^d$ and $Z^s$ are uncorrelated and centered.  From \eqref{eqn:Wright}, we can solve for the equilibrium $P$ and $Y$ in terms of $U^d$ and $U^s$:
$$
P  =  \frac{1}{\beta_1 - \alpha_1} (U^d - U^s)\quad\text{and}\quad
Y  =  \frac{\alpha_1}{\beta_1 - \alpha_1} (U^d - U^s) + U^d.
$$
From \eqref{eqn:U}, a unit increase in $Z^s$ moves $U^s$ up by $\beta_2$.  From the previous expressions, a unit increase in $U^s$ results in a shift of the equilibrium price $P$ of $-(\beta_1 - \alpha_1)^{-1}$ and a shift of the equilibrium quantity $Y$ by  $-(\beta_1 - \alpha_1)^{-1}\alpha_1$.  Therefore, the overall effect of a unit increase in $Z^s$ is a shift in $P$ of  $-(\beta_1 - \alpha_1)^{-1}\beta_2$ and a shift  in $Y$ of $-(\beta_1 - \alpha_1)^{-1}\alpha_1 \beta_2$. The elasticity $\alpha_1$ is identified by the ratio of these two shifts. In turn, the two shifts can be identified as coefficients of the linear projections of $P$ on $Z^s$ and $Y$ on $Z^s$:
\begin{eqnarray*}
P & = &  - \frac{\beta_2 }{\beta_1 - \alpha_1} Z^s + V_P, \quad V_P \perp Z^s;\quad\text{and}\\
Y &=&  - \frac{\alpha_1 \beta_2 }{\beta_1 - \alpha_1}  Z^s + V_Y, \quad V_Y \perp Z^s.
\end{eqnarray*}
Then, taking the ratio  of the two coefficients gives the same formula as the previous method of moments approach. This second proposal of Wright is the `indirect least squares' method. Wright used this reasoning together with a plot like Figure \ref{fig:DAG2} to support this reasoning, inspired by his son's work on causal diagrams.  

This method cleverly bypasses the fact that the latent nodes $U^d$ and $U^s$ are unobserved. Had we observed them, we could have used Sewall Wright's causal path diagrams (linear DAGs) where all the parameters can be learned by linear regression, and we could then deduce the values of the elasticities from these parameters. This would still not be a trivial exercise, which further underscores the depth of Wright's analysis.

\section{Counterfactuals under Tax Interventions}

Wright motivated the importance of the structural model \eqref{eqn:Wright} with an analysis of hypothetical policies such as the imposition of a tariff on imports. Here, we consider a simplified setting where all producers are foreign and consumers are domestic.\footnote{Wright's Appendix B considers a more general setting where there are both domestic and foreign consumers and producers.} We also work with a log-linear system, whereas Wright used a linear system.  We analyze the effect of introducing a tariff on imports proportional to the price, which is levied on the producers. This tariff creates a gap between the price that the consumers pay and the price that the producers receive. Thus, if the consumers pay a (log) price $p$, then the producers receive a (log) price $p + \log(1-\tau) \approx p - \tau$, where $\tau>0$ is the tariff rate.

To analyze the effect of the tariff, it is convenient to introduce the counterfactual demand and supply equations resulting from the policy. The counterfactual demand is the same as the original demand, whereas the counterfactual supply is shifted. In particular, if the denote by $D^\star(p)$ and $S^\star(p)$ the counterfactual demand and supply, respectively, then
\begin{equation} \label{eqn:counter} 
  \begin{aligned} 
   && D^\star(p) &= D(p) = \alpha_1 p + U^d\quad\text{and}&&\\
    && S^\star(p) &= S (p-\tau) = \beta_1 (p - \tau) + U^s, && 
\end{aligned}
\end{equation}
assuming that the policy does not affect the stochastic factors $U^d$ and $U^s$. The counterfactual equilibrium variables $(P^\star, Y^\star)$ obtained from the condition $D^\star(P^\star)  = S^\star(P^\star) =Y^\star$ in \eqref{eqn:counter} are
$$
P^\star =  P + \Delta P\quad\text{and}\quad Y^\star  =   Y + \Delta Y,
$$
where
$$
\Delta P := c \tau \geq 0\quad\text{and}\quad \Delta Y :=  \alpha_1 \Delta P \leq 0
$$
under the standard assumption that the elasticities $\alpha_1 \leq 0$ and $\beta_1 \geq 0$, so that $c :=  \beta_1/\left(\beta_1 - \alpha_1\right) \geq 0$. That is, introducing the tariff increases the equilibrium price by $\Delta P$ and reduces the equilibrium quantity by $|\Delta Y|$, where the magnitude of the changes depends on the demand and supply elasticities and the tariff rate. Figure \ref{fig:CDAG} provides a causal diagram behind \eqref{eqn:counter} based on the general setup of Figure \ref{fig:DAG}. Here the counterfactual policy replaces the supply node $S(\cdot)$ by the counterfactual supply node $S^\star (\cdot)$.\footnote{\label{fn:dofix}Interestingly, this does not immediately correspond to any standard do-interventions or fix-interventions in the modern identification work on DAGs ({\em e.g.} Pearl, 2009a; Richardson and Robins, 2014; and Heckman and Pinto, 2015), because the intervened node preserves incoming and outgoing edges. To put this intervention in the do framework, we need to introduce another node $\tau$ and the edge $\tau \mapsto S^\star (\cdot) = S(\cdot - \tau)$. Here the effect $\tau \mapsto S(\cdot - \tau)$ is identified from theoretical introspection, a `thought experiment', rather than from any empirical observation. See Pearl, J. (2009a). Causal inference in statistics: an overview. {\em Statistics Surveys 3}, 96--146; Heckman, J. J. and R. Pinto (2015). Causal analysis after Haavelmo. {\em Econometric Theory 31}(1), 115--151; and Footnote \ref{fn:finecycles} for a reference to Richardson and Robins (2014).}

\begin{figure}[t]
\vspace*{10pt}
\begin{center}
\begin {tikzpicture}[-latex, auto, node distance =1.5cm and 3cm, on grid, thick, 
  observed/.style ={circle, top color=white, bottom color = blue, draw, black, text=black , minimum width =1.75 cm},
  unobserved/.style ={circle, top color=white, bottom color = yellow, draw, black, text=black , minimum width =1.75 cm}, 
  eqb/.style ={rectangle, top color=white, bottom color = magenta, draw, black, text=black , minimum width =1.75 cm}, 
empty/.style ={circle, top color=white, bottom color = white, draw,  white, text= white, minimum width =.5 cm}]

\node[unobserved]  (Yd) at (2,2) {$D(\bullet)$};
\node[unobserved]   (Ys) at (2,-2) {$S^\star(\bullet)$};
\node[observed]   (Zs) at (-2,-2) {$Z^s$};
\node[observed]   (Zd) at (-2,2) {$Z^d$};
\node[observed]    (P) at (6,2) {$P^\star$};
\node[observed]    (Y) at (6,-2) {$Y^\star$};
\path (Ys) edge (Y);
\path (Yd) edge (Y);
\path (Ys) edge (P);
\path (Yd) edge (P);
\path (Zd) edge (Yd);
\path (Zs) edge (Ys);
\node[unobserved]  (L2) at (0,0)  {$\cstoDS$};
\node[unobserved]  (L1) at (-4,0)  {$\cstoZ$};
\path (L1) edge (Zd);
\path (L1) edge (Zs);
\path (L2) edge (Yd);
\path (L2) edge (Ys);\end{tikzpicture} 
\end{center}
{\footnotesize \textbf{Note:} Same notation as in Figure \ref{fig:DAG}. The policy replaces the supply node $S(\cdot)$ by the counterfactual supply node $S^\star (\cdot)$.  See Footnote \ref{fn:dofix} for an alternative graphical representation of this intervention.}

\caption{A causal diagram of Wright's counterfactual model.\label{fig:CDAG}} 
\end{figure}
 
Wright worked further to determine optimal tariff rates for taxing imports.\footnote{He worked with linear demand and supply systems, whereas our exposition proceeds with log-linear forms, which seems more reasonable. The point of the discussion that follows is to illustrate the use of elasticities for welfare calculations. } Roughly speaking, imposing tariffs reduces consumer surplus due to increased prices and decreased quantities. At the same time, the government can pay back  consumers the collected tariffs in the form of a rebate or spend it, at best, efficiently on public goods.\footnote{Of course, globally, this entails a deadweight loss since the reduction in consumer and producer surpluses cannot be offset by the tariff revenue collected.}  For small $\tau$, the change in consumer surplus divided by the base revenue $\exp(Y+P)$ is approximately\footnote{Let  $\mP := e^P$, $\mQ := e^Y$, $\mP^\star := e^{P^\star}$, $\mQ^\star := e^{Y^\star}$, $\mathrm{Rev} := \mP \mQ$, and $\mathrm{Rev}^\star :=  \mP^\star \mQ^\star$.
The change in consumer surplus, using the trapezoid approximation, is
$$
\Delta CS \approx \frac{\mQ + \mQ^\star}{2} (\mP - \mP^\star) = - \left ( 1+ \frac{\Delta Y}{2} \right) \Delta P \mathrm{Rev},
$$
and the total tariff revenue collected is
$$
\tau \mathrm{Rev}^\star = 
\left( \tau + \tau \left ( 
\frac{\mathrm{Rev}^\star - \mathrm{Rev} }{\mathrm{Rev}}
\right ) \right ) \mathrm{Rev},
$$
where
$$
\frac{\mathrm{Rev}^\star - \mathrm{Rev} }{\mathrm{Rev}} \approx \Delta P + \Delta Y + \Delta P \Delta Y.
$$
}
$$
-\left ( 1+ \frac{\Delta Y}{2} \right) \Delta P = - c \tau - \frac{\alpha_1 c^2 }{2}\tau^2.  
$$
and the total tariff revenue collected divided by the base revenue  $\exp(Y+P)$ is approximately
$$
\tau + (\Delta Y + \Delta P + \Delta Y \Delta P)\tau = \tau + c (1 + \alpha_1) \tau^2 
 +   c^3 \alpha_1^2 \tau^3.
$$
We can compute an approximate optimal tariff level by plotting the sum of the two expressions (including quadratic and possibly cubic terms) against a small range of potential values of $\tau$ and choose that $\tau$ where this sum is largest.\footnote{If domestic consumers do receive the rebate or the benefits of the increased public good provision, then changing $\tau$ from zero to a small positive value increases domestic consumer welfare by $-c \tau + \tau>0$ if  $c = (\beta_1 - \alpha_1)^{-1} \beta_1 < 1.$ This is always the case unless the supply is perfectly elastic.} 

\section{Analysis of Wright's Model via the Generalized Method of Moments}
%===================================================================================

In what follows it will be convenient to partial out some variables.  To set this up, we define a partialing-out operator that acts on any square integrable variable $V$ subtracting the best linear predictor $L (V \mid M) := M'(\Ep MM')^{-}\Ep MV$ of $V$ on $M$, and creating the population residual:\footnote{For a matrix $A$, $A^-$ denotes the Moore-Penrose inverse.}
\begin{equation}\label{eq:partialing-out}
\tilde V_1 := V - L (V \mid W, Z^d)\quad\text{and}\quad\tilde V_2 := V - L (V \mid W, Z^s).
\end{equation}
Evidently the partialing-out operator is linear, because the best linear predictor is a linear operator; that is, for any $a,b \in \mathbb{R}$,
$$
L (aV_1 + bV_2 \mid M) = a L (V_1\mid M) + b L (V_2\mid M).
$$
Hence, applying \eqref{eq:partialing-out} to partial out $W,Z^d$ from the first equation in \eqref{eqn:SEM} and $W,Z^s$ from the second equation in \eqref{eqn:SEM} gives the simultaneous equations model
\begin{equation}\label{eqn:SEM3}
\begin{aligned}
&&  \tilde Y_1 - \alpha_1 \tilde P_1 = \epsilon^d, \quad &\epsilon^d \perp \tilde Z^s_1;\quad\text{and} && \\
&&  \tilde Y_2 - \beta_1 \tilde  P_2  =   \epsilon^s, \quad &\epsilon^s \perp \tilde Z^d_2. &&
\end{aligned}
\end{equation}
This corresponds to the moment restrictions
\begin{equation}\label{key2}
\ \Ep ( \tilde Y_1 - \alpha_1  \tilde P_1) \tilde Z^s_1 = 0\quad\text{and}\quad\Ep ( \tilde Y_2 - \beta_1  \tilde P_2) \tilde Z^d_2 = 0.
 \end{equation}
The two equations have unique solutions if $\Ep \tilde P_1 \tilde Z^s_1$ and  $\Ep \tilde P_2 \tilde Z^d_2$ have full column rank, that is, if the residualized prices and instruments are correlated. Therefore, the partialing-out approach allows us to very clearly  see how the identification can be achieved here. It is also convenient for inference, especially when $W$ is high-dimensional. For now we assume that the residuals above are known, but we can replace them by empirical versions in the definition of the generalized method of moments (GMM) estimator, without affecting its large sample properties. We explain this further below.

In order to develop the approach systematically, we need to set up some notation. Let $ \theta_0 := (\alpha_1, \beta_1)'$ denote the value of the parameter vector that satisfies \eqref{key2}, and let $\theta := (a, b)'$ denote any potential value that the parameter vector can take in a parameter space $\Theta \subset \Bbb{R}^2$. 
Denote $X:=(\tilde Y_1,\tilde P_1,\tilde Z^s_1,\tilde Y_2,\tilde P_2,\tilde Z^d_2)$ and define the `score' function:
\begin{equation}\label{eq:GMMscore}
g(X, \theta): = \left [ \begin{array}{c}
g_1(X, a)  \\
 g_2(X, b)
 \end{array} \right] :=  \left [ \begin{array}{c}
(\tilde Y_1- a \tilde P_1)  \tilde Z^s_1 \\
 (\tilde Y_2- b \tilde P_2) \tilde Z^d_2
 \end{array} \right].
 \end{equation}
Let   $g(\theta) : =  \Ep g(X, \theta)$ be the average score or moment function. The score and average score are affine:
$$
g(X, \theta) = g(X, 0) + G(X) \theta\quad\text{and}\quad g(\theta) = g(0) + G \theta, 
$$
where
$$
G(X) := \left[ \begin{array}{cc}
G_1(X) & 0 \\
0  & G_2 (X) \end{array} \right] 
= \left[ \begin{array}{cc}
 -   \tilde Z^s_1 \tilde P_1 & 0 \\
0  & -  \tilde Z^d_2  \tilde P_2 \end{array}\right]
$$
and
$$
G :=  \left[ \begin{array}{cc}
 G_1 & 0 \\
0  & G_2 \end{array} \right]
= 
\left[ \begin{array}{cc}
 \Ep G_1(X) & 0 \\
0  & \Ep G_2(X) \end{array} \right].
$$

The true parameter value $\theta_0 \in \Theta$ satisfies $g(\theta_0) = 0$. It is identified by these linear equations if the Jacobian matrix $G$ is of full column rank. Alternatively, under this rank condition, $\theta_0$ can be characterized as
$$
\theta_0 = \arg\min_{\theta \in \Theta } g(\theta)'A g(\theta),
$$
for any symmetric positive definite matrix $A$, or as the solution to the corresponding first order conditions $0 = G' A  g(\theta_0) = G' A ( g(0) +  G \theta_0)$. Since $\theta\mapsto g(\theta)$ is affine, this is a quadratic optimization problem with solution $\theta_0 = M  g(0)$, where
\begin{equation}\label{explicitsolution}
M := - (G' A G)^{-1} G' A
\end{equation}
is well-defined because $G$ is of full column rank.  This characterization applies only if $\theta_0$ is in the interior of $\Theta$.

To estimate $\theta_0$, we have data $\{X_i\}_{i=1}^n$, which are identical copies of $X$, and, as a leading case we assume  that they are also independent (i.i.d.).  We form the empirical moment function
$$
\hat g(\theta) := \En g(X, \theta),
$$
where $\En$ denotes the empirical average. 
Then, we can construct a GMM estimator of $\theta_0$ as 
\begin{equation}
 \label{eq:GMM}
\hat \theta \in \arg\min_{\theta \in \Theta}  \  \hat g(\theta) '\hat A \hat g(\theta),
\end{equation}
where $\hat A$ is a symmetric positive-definite matrix, possibly data-dependent,  that converges to a nonstochastic symmetric positive-definite matrix $A$; that is, $\hat A \to_P A$.  This GMM estimator has an analytical closed form solution $\hat \theta = \hat M  \hat g(0)$, where
\begin{equation}\label{explicit}
\hat M := - (\hat G' \hat A \hat G)^{-1} \hat G' \hat A,
\end{equation}
 $\hat g(0) := \En g(X,0)$, and $\hat G := \En G(X)$, provided that $\hat M$ is well defined and the solution is in the interior of the parameter space. This follows by solving the first-order conditions for the minimization in \eqref{eq:GMM} that defines the GMM estimator: 
$$0 = \hat G' \hat A \hat g(\hat \theta) = \hat G' \hat A ( \hat g(0) +  \hat G\hat \theta).$$ 

An important practical and theoretical matter is the choice of the weighting matrix $A$.  The optimal choice, which minimizes the asymptotic variance of the  GMM estimator, is
$$
A = \Omega^{-1},    \quad \Omega :=  \Var ( \sqrt{n} \hat g(\theta_0)).
$$
In the case of i.i.d. data,  ${ \Omega = \Var g(X, \theta_0) = \Ep g(X, \theta_0) g(X, \theta_0)'.  }$
Note that $\Omega$ is unknown, and for this reason, we often use  the following algorithm to compute the GMM estimator. Let $K \geq 2$ be a pre-specified number of steps.

\begin{algorithm}[Iterative GMM]\label{alg:gmm} \hfill \break \vspace{-.75cm}
\begin{itemize}
\item[1.]  Set $k=1$, $\hat A = \hat \Omega^{-1}$, for block diagonal $\hat \Omega$ with top-left block $\En \tilde Z^s_1 \tilde Z^{s\prime}_1$ and bottom-right block $\En \tilde Z^d_2 \tilde Z^{d\prime}_2$, and obtain $\hat \theta_0$ using \eqref{eq:GMM}.\\
\item[2.] Set $\hat A = \hat \Omega^{-1}$, for $\hat \Omega = \widehat{\Var} (\sqrt n  \hat g(\theta))  |_{\theta = \hat \theta_{k-1}}$, and obtain $\hat \theta_k$ using \eqref{eq:GMM}.\\
\item[3.] Repeat the previous step for $k = 2, \ldots, K$. Report $\hat \theta_K$ as the estimator of $\theta_0$.
\end{itemize}
\end{algorithm}
Note that Step 1 returns the conventional two-stage least squares estimators.  In step 2, we need to specify an estimator of $\Omega$;  under i.i.d. sampling we can use
$$
\widehat{\Var} (\sqrt n  \hat g(\theta)) = \Vn g(X_i, \theta) := \En g(X_i, \theta) g(X_i, \theta)' - 
\En g(X_i, \theta) \En g(X_i, \theta)'.
$$
For dependent data, we can use the Newey-West variance estimator. Two steps ($K=2$) are sufficient to reach full efficiency, although at least one additional step might be desirable to obtain a more efficient estimator of the variance and  improve the finite-sample properties of the estimator.\footnote{The common practice is to use $\En g(X_i, \theta) g(X_i, \theta)'$  instead of  $\Vn g(X_i, \theta)$, thought the latter seems to be better behaved in practice.}

Using the notation defined above we can conclude that
$$
\sqrt{n} (\hat \theta - \theta_0) = \hat M \sqrt{n}\hat g(\theta_0).
$$
Under strong identification, namely that the minimal eigenvalue of $G'AG$ is bounded away from zero,
and other mild regularity conditions, the GMM estimator is approximately linear and Gaussian:
\begin{equation}\label{eq:gmm_ad}
\sqrt{n} (\hat \theta - \theta_0)
= M \sqrt{n}\hat g(\theta_0) +o_P(1)    \ad   M \ \mathcal{N}(0, \Omega)= \mathcal{N}(0,V_A), \quad V_A := M\Omega M',
\end{equation}
where $M$ is defined in \eqref{explicitsolution} and 
$\sqrt{n}\hat g(\theta_0) \ad \mathcal{N}(0, \Omega)$. If the efficient weighting matrix is used $A =  \Omega^{-1}$, then variance simplifies to be:
$$V_{A} = V_{\Omega^{-1}} = (G'\Omega^{-1} G)^{-1}.$$
To operationalize this result for inference, we need to find a consistent estimator of $V_A$. We can do so employing the standard plug-in principle.

Hansen, Heaton, and Yaron (1996) introduced an alternative version of the GMM estimator called the continuous-updating estimator (CUE).\footnote{Hansen, L. P., J. Heaton and A. Yaron (1996). Finite sample properties of some alternative GMM estimators. {\em Journal of Business and Economic Statistics 14}(3), 262--80.} The CUE is
$$
\hat \theta^* \in \arg\min_{\theta \in \Theta} \hat g(\theta)' \hat A (\theta) \hat g(\theta), \quad \hat A(\theta) = \hat \Omega(\theta)^{-1},
$$
where
$
\hat \Omega(\theta)
$
is a consistent estimator of $\Omega(\theta) := \Var(\sqrt{n} \hat g(\theta))$.
This form is quite intuitive because it uses inverse of a variance matrix (indexed by $\theta$) directly in the formulation of the GMM estimator.  Unlike the iterative GMM estimator, CUE avoids iteration by obtaining the  estimator of the variance matrix simultaneously  with the estimator of $\theta$ (`updating continuously').  Under i.i.d. sampling, 
$$
\hat \Omega(\theta) =  \Vn g(X, \theta).
$$
Under time series sampling, we can use a Newey-West-type  estimator.  The CUE has the same properties as efficient GMM under strong identification and suitable regularity conditions, 
$$
\sqrt{n} (\hat \theta^* - \hat \theta) \to_P 0,
$$
but its objective function is convenient to perform inference under weak identification (see Section \ref{ss:weak}).

\subsection{Limited information versus full information GMM}
%--------------------------------------------------------------------------------------------------------------------------------------------------
We can estimate Wright's model using  two approaches:
\begin{enumerate}
\item In the `full information' or `system' approach, we estimate $\alpha_1$ and $\beta_1$ jointly via GMM, employing a (jointly) optimal weighting matrix.\\
\item In the `limited information' or `equation-by-equation' approach, we estimate $\alpha_1$ and $\beta_1$ separately, each with its own GMM estimator based on each block of equations. We can artificially view this approach  as a joint GMM with a block-diagonal weighting matrix. 
\end{enumerate}
The  full information approach is generally more efficient if some regularity conditions, particularly strong identification conditions, hold for the joint estimation problem.  The limited information approach is generally more robust and more widely applicable, and this is the approach Wright used.

For the purpose of statistical analysis, we can treat both approaches in the GMM framework. Both use 
$$
g(X, \theta) = [g_1(X, a)', g_2(X, b)']',
$$
defined in \eqref{eq:GMMscore}, for the `demand' and `supply' scores. The full information approach employs the general optimal weighting matrix
$$
A = \left [ \Var g(X, \theta_0) \right]^{-1},
$$
whereas the limited information approach employs the block-diagonal weighting matrix
$$
A = \left [\begin{array}{cc}
\Var g_1(X, \alpha) &  0 \\
0 &  \Var g_2(X, \beta) \end{array}  \right]^{-1}.
$$
Both matrices need to be estimated using an interative procedure like Algorithm \ref{alg:gmm}.

 \subsection{Weak identification}
\label{ss:weak}
%--------------------------------------------------------------------------------------------------------------------------------------------------
Weak identification arises when the minimal eigenvalue of $G'AG$ is close to zero--- that is, when $G$ is close to being column rank deficient--- relative to the sampling error. This occurs when either $Z^s$ or $Z^d$ is a weak instrument for $P$. Under weak identification, the Gaussian distribution in \eqref{eq:gmm_ad} may provide a poor approximation to the finite-sample distribution of the GMM estimator and the estimator itself starts to behave erratically, because we are dividing by a random quantity that fluctuates dangerously  close to zero; see the expression for $\hat M$  in \eqref{explicit}.

\begin{remark}[Weak instrument asymptotics]
Staiger and Stock (1997) developed an asymptotic approximation to the weak identification case where the data streams $\{X_{i,n}\}_{i=1}^\infty$ are identically distributed random vectors with a law $F_n$ that changes with $n$.\footnote{Staiger, D. and J. H. Stock (1997). Instrumental variables regression with weak instruments. {\em Econometrica 65}(3), 557--86.}  This is an alternative to the conventional asymptotic approximation where  $n \to \infty$  but $F$ is fixed. In the `weak instrument asymptotics', as $n \to \infty$, $F_n$ is such that  the minimal eigenvalue of $G'AG= G_n'AG_n$ is zero or drifts to zero as $n \to \infty$, where $G_n= (\partial/\partial \theta') \Ep g(X_{i,n}, \theta_0)$ fast enough, so that the conventional Gaussian approximation in \eqref{eq:gmm_ad} breaks down. However, it is still possible to make inference using ideas due to Anderson and Rubin (1949) and Stock and Wright (2000), and related ideas described below.\footnote{See Anderson, T. W. and H. Rubin (1949). Estimation of the parameters of a single equation in a complete system of stochastic equations. {\em The Annals of Mathematical Statistics 20}(1), 46--63; and Stock, J. H. and J. H. Wright (2000). GMM with weak identification. {\em Econometrica 68}(5), 1055--96.}
\end{remark}

The Anderson-Rubin (AR) statistic is the (scaled) objective function of the CUE:
$$
S(\theta) :=   n \hat g(\theta)'\hat A(\theta) \hat g(\theta). 
$$
Suppose that the empirical moments are asymptotically normally distributed,  
$$\sqrt{n}(\hat g(\theta) - g(\theta)) \ad \mathcal{N}(0, \Omega(\theta)),$$
for each $\theta \in \Theta$. Assume that $\theta \mapsto \hat A(\theta)$
obeys $\hat A(\theta) \to_P \Omega(\theta)^{-1}$ for each $\theta \in \Theta$.  
Then, for any $\theta_0$ such that $g(\theta_0) = 0$, 
$$
S(\theta_0) =   n \hat g(\theta_0)'\hat A(\theta_0) \hat g(\theta_0) \ad \chi^2(m),
$$
where $m$ is the dimension of $g(\theta)$. 
Consequently,  the confidence region 
$$
CR_{1-p} := \{ \theta \in \Theta:  S(\theta) \leq c_{1-p}\},
$$ 
where $c_{1-p}$
is the $(1-p)$-quantile of $\chi^2 (m)$, contains $\theta_0$ with asymptotic probability $1-p$:
$$
\Pr\left ( \theta_0 \in CR_{1-p} \right )  = \Pr( S(\theta_0) \leq c_{1-p}) \to  \Pr( \chi^2(m) \leq c_{1-p})  = 1-p, \quad n \to \infty.
$$ 
This confidence region is robust to weak identification because its validity does not depend on the properties of the eigenvalues of $G'AG$. 
We can approximate this confidence region in practice by specifying a grid of parameter values and then collecting all values $\theta$ on this grid such that the AR statistic $S(\theta)$ is less than the critical value.  

The Anderson-Rubin approach can be  conservative when there are more moment conditions than parameters. We may want to look at the deviation of the AR statistic $S(\theta)$ from its minimal value over the parameter set instead, which gives the quasi likelihood ratio (LR) statistic  
$$
LR(\theta) := S(\theta) - \inf_{\bar \theta \in \Theta} S(\bar \theta).
$$
High values of $LR(\theta)$ provide evidence against the plausibility of an hypothesized $\theta$.

The distribution of this statistic is nonstandard, but one may pursue a conditional approach. Intuitively we  need to condition on how strong the first stage $\hat G$ is--- that is, how far $\hat G$ is from being column rank deficient--- in this inferential procedure, but we need to orthogonalize $\hat g(\theta)$ with $\hat G$.  Moreira (2003) and Andrews and Mikusheva (2016) have proposed to do so as follows.\footnote{\label{fn:LR}See Moreira, M. J. (2003). A conditional likelihood ratio test for structural models. {\em Econometrica 71}(4), 1027--48; and Andrews, I. and A. Mikusheva (2016). Conditional inference with a functional nuisance
parameter. {\em Econometrica 84}(4), 1571–612.} Let
$(\theta, \bar \theta)\mapsto\widehat{\Omega}(\theta, \bar \theta)$ be such that  $\widehat{\Omega}(\theta, \bar \theta)\to_P \mathrm{Cov}(\sqrt{n} \hat g(\theta), \sqrt{n} \hat g(\bar \theta) )$ for all $(\theta,\bar\theta)\in\Theta^2$.  For $\theta_0$ such that $g(\theta_0) = 0$, 
introduce the conditioning statistic
$$
\hat h(\theta,\theta_0) :=\hat g (\theta)-\widehat{\Omega}\left(\theta, \theta_0 \right) \widehat{\Omega}\left(\theta_0, \theta_0\right)^{-1} \hat g\left(\theta_0 \right).
$$
Let
$$
\hat g^*(\theta,\theta_0):=\hat h(\theta,\theta_0)+\widehat{\Omega}\left(\theta, \theta_0 \right) \widehat{\Omega}\left(\theta_0, \theta_0 \right)^{-1} \xi(\theta_0)/\sqrt{n},
$$
where $\xi(\theta_0) \sim N(0, \widehat{\Omega}(\theta_0, \theta_0))$. Then,
$$
LR(\theta_0) \mid \hat h(\cdot,\theta_0) \ad  LR^* (\theta_0),
$$
where
\begin{align*}
LR^* (\theta_0) := &\, n \hat g^*(\theta_0,\theta_0)' \widehat{\Omega}(\theta_0, \theta_0)^{-1}  \hat g^*(\theta_0,\theta_0) - n \inf_{\theta \in \Theta} \hat g^*(\theta,\theta_0)' \widehat{\Omega}(\theta,\theta)^{-1}  \hat g^*(\theta,\theta_0)\\
 = &\, \xi(\theta_0)' \widehat{\Omega} (\theta_0, \theta_0)^{-1}\xi(\theta_0) -n \inf_{\theta \in \Theta} \hat g^*(\theta,\theta_0)' \widehat{\Omega}(\theta,\theta)^{-1}  \hat g^*(\theta,\theta_0).
\end{align*}
The critical value  $c_{1-p}(\theta_0;\hat h(\cdot,\theta_0),\widehat{\Omega}(\cdot,\cdot))$ is obtained by simulation as the $1-p$ quantile of $LR^*(\theta_0)$.  The simulation keeps $\hat h(\cdot,\theta_0)$ and $\widehat{\Omega}(\cdot,\cdot)$ fixed, creates many independent draws of $\xi(\theta_0)$ from $N(0, \widehat{\Omega}(\theta_0, \theta_0))$, and then uses all three to compute values of $LR^*(\theta_0)$.  The asymptotic confidence region of level $1-p$ is obtained by inverting the test:
$$
CR_{1-p}^* := \{ \theta \in \Theta: LR(\theta) \leq  c_{1-p}(\theta;\hat h(\cdot,\theta),\widehat{\Omega}(\cdot,\cdot))\}.
$$
 It is also possible to consider a similar inferential approach using Wald statistics, but LR seems to perform better in many examples. See Moreira (2003) and Andrews and Mikusheva (2016) for further details.

 \subsection{Partialing out, high-dimensional $W$, and machine learning} 
 %--------------------------------------------------------------------------------------------------------------------------------------------------
The partialing out of $W$ in \eqref{eq:partialing-out} uses population moments, and is not feasible in finite samples.  To operationalize the approach in finite samples,  if $W$ is low-dimensional, we can use least squares to partial out $W$ from all other variables.\footnote{Alternatively we can simply use GMM with all parameters being estimated jointly.} If $W$ is high-dimensional, we can use machine learning procedures such as the least absolute shrinkage and selection operator (LASSO) to partial out $W$ from other variables in the data and apply the methods described above.  Specifically, in the definition of the GMM estimator, instead of the population residuals in \eqref{eq:partialing-out}, we would use sample residuals
 $$
\check V_1 := V_1 - \hat L (V_1 \mid W, Z^d)\quad\text{and}\quad 
\check V_2 := V_2 - \hat L (V_2 \mid W, Z^s),
$$
 where $\hat L$ is estimated via LASSO regression.  For estimation purposes, the modified score function then becomes
$$
\check g(X, \theta): = \left [ \begin{array}{c}
\check g_1(X, a)  \\
 \check g_2(X, b)
 \end{array} \right] :=  \left [ \begin{array}{c}
(\check Y_1- a \check P_1)  \check Z^s_1 \\
 (\check Y_2- b \check P_2) \check Z^d_2.
 \end{array} \right].
$$
By arguments as in Chernozhukov, Hansen, and Spindler (2015a,b), this score function is asymptotically equivalent to the score function $g(X, \theta)$; that is
 $$
\sup_{\theta \in K}  \Big | \sqrt{n} ( \En g(X, \theta) - \En \check g(X, \theta))  \Big | \to_P 0
 $$
 for any compact set $K$,   provided that $L (V_1 \mid W, Z^d)$ and $L (V_2 \mid W, Z^s)$  admit sufficiently good sparse approximations with respect to $W$.\footnote{See Chernozhukov, V., C. Hansen and M. Spindler (2015a). Post-selection and post-regularization inference in linear models with very many controls and instruments. {\em American Economic Review: Papers and Proceedings 105}(5), 486--90; and Chernozhukov, V., C. Hansen and M. Spindler (2015b). Valid post-selection and post-regularization inference: an elementary, general approach. {\em Annual Review of Economics 7}(1), 649--88.} Therefore, all the inference properties apply to $\check g$ in place of the infeasible $g$. This property is a consequence of the Neyman orthogonality property with respect to the parameters of the best linear projections in the partialing-out steps.  For machine learning beyond LASSO, we need to use cross-fitting to eliminate biases that result from overfitting (see Chernozhukov {\em et al.}, 2018, for the analysis of partially linear instrumental regression models).\footnote{Chernozhukov, V., D. Chetverikov, M. Demirer, E. Duflo, C. Hansen, W. Newey and J. Robins (2018). Double/debiased machine learning for treatment and structural parameters. {\em The Econometrics Journal 21}(1), C1--C68.}

 \section{Some Connections to Contemporary Work}
 \label{s:contemporary}
 %===================================================================================

\subsection{Structural causal models and DAGs} 
%--------------------------------------------------------------------------------------------------------------------------------------------------
Judea Pearl's work on causal DAGs and structural causal models is quintessential in modern causal inference. He has made several fundamental contributions.   One key contribution is to formally define causal DAGs as Markov networks representing the joint distribution of variables and factorization of distributions according to the DAG. The factorization implied by the DAG in Figure \ref{fig:DAG}, augmented with $W$ as laid out in the note to this figure, is
\begin{align*}
&f(y, p, d, s, k_2, z^d, z^s, w,k_1)\\ 
&~~~~~~~~= 
f(y| d, s) f(p| d, s) f(d | z^d,w,k_2) f(s| z^s,w,k_2) f(k_2)
f(z^d|w,k_1)f(z^s|w,k_1) f(w,k_1),
\end{align*}
where $f$ denotes the probability measure $\mathrm{d}\Pr$ of the random element $$(Y, P, D(\cdot), S(\cdot), \cstoDS, Z^d, Z^s,W,\cstoZ)$$ and $(y, p, d, s, k_2,z^d, z^s, w,k_1)$ denotes values of this random element.\footnote{ That is, an event $S$ is assigned probability  $\int_S \mathrm{d}\Pr = \int \mathds{1}_S(y, p, d, s, k_2,z^d, z^s, w,k_1) f (y, p, d, s, k_2,z^d, z^s, w,k_1)$.}   The factorization encodes the order in which blocks of variables get determined given parental nodes. It also encodes all the conditional independencies in the model. The formula above is rather bulky, however, and it is much nicer to `visualize' the formula through the DAG itself in Figure \ref{fig:DAG}.

However, the representation of the distribution via the DAG by itself does not have any causal content. To endow DAGs with causal interpretation, Pearl (1995, 2009b) links the statistical model to a system of structural equations, building upon Haavelmo, which are said to represent invariant relationships that retain autonomy under interventions.\footnote{See Pearl, J. (1995). Causal diagrams for empirical research. {\em Biometrika 82}(4), 669--88; and the reference to Pearl (2009b) in Footnote \ref{fn:cyclical}.} For example, in our context, the structural equations are those that we laid out. The autonomy corresponds to the assumption that the tariff intervention on the supply node does not affect the nondescendant nodes: the supply and demand shifters and the demand itself. This autonomy is in line with Wright's reasoning that we presented earlier.

Another key contribution of Pearl (2009b) is the method for determining conditional independence (or exclusion restrictions) from the graph structure.  In our context, the key is to verify independence conditions (exclusion restrictions) such as
\begin{equation}
\label{eq:ci}
\left(D(p)\right)_{p\in\mathbb{R}} \indep Z^s  \mid Z^d, W,
\end{equation}
which states that the demand curve is independent of the supply shifters $Z^s$, conditional on the demand shifters $Z^d$ and common shifters $W$.  To illustrate how graphical criteria can be used to verify conditional independence restrictions like \eqref{eq:ci}, consider the simplified DAG in Figure \ref{fig:DAG} (we ignore $W$ in this illustration). On the path 
\begin{equation}
\label{eq:path}
Z^s \longleftarrow \cstoZ \longrightarrow Z^d  \longrightarrow D(\cdot),
\end{equation}
the latent factor $\cstoZ$ creates statistical dependence between the supply and demand shifters.  It also generates dependence between the supply shifter $Z^s$ and demand $D(\cdot)$. We can remove this dependence by conditioning on $Z^d$, as in \eqref{eq:ci}.  We can graphically verify this exclusion restriction by checking that $Z^s$ and $D(\cdot)$ are `d-separated' by $Z^d$.\footnote{D-separation is a graphical criterion used to determine whether two variables are independent of each other conditional on others. See Verma, T. and J. Pearl (1988). Influence diagrams and d-separation. Technical Report CSD 880052, Cognitive Systems Laboratory, UCLA.}  This requires that all paths between $Z^s$ and $D(\cdot)$ are d-separated  by $Z^d$. The path in \eqref{eq:path} is d-separated by $Z^d$ because it contains the chain $\cstoZ \rightarrow Z^d  \rightarrow D(\cdot)$, so that fixing $Z^d$ would `block' the effect of $\cstoZ$ on $D(\cdot)$ on this path. The other three paths between $Z^s$ and $D(\cdot)$,
 \begin{align}
 \label{eq:collider}
 &Z^s \longrightarrow S(\cdot) \longleftarrow\cstoDS \longrightarrow D(\cdot),\nonumber\\
 &Z^s \longrightarrow S(\cdot) \longrightarrow P \longleftarrow D(\cdot),\text{ and}\\
 &Z^s \longrightarrow S(\cdot) \longrightarrow Y \longleftarrow D(\cdot),\nonumber
 \end{align}
all contain a `collider', a node with two incoming arrows ($S(\cdot)$, $P$, and $Y$, respectively). By themselves, such paths do not create dependence between their terminal nodes, $Z^s$ and $D(\cdot)$. However,  conditioning on the collider on the path, or any of its descendants, would generate dependence between the nodes pointing to the collider. For example, conditioning on $S(\cdot)$ would make its causes $Z^s$ and $\cstoDS$ and thus $Z^s$ and $D(\cdot)$ dependent. However, none of the colliders is $Z^d$ or one of its descendants, so each of the three paths in \eqref{eq:collider} are d-separated by $Z^d$. As all four paths between $Z^s$ and $D(\cdot)$ are d-separated by $Z^d$,  $Z^s$ and $D(\cdot)$ are d-separated by $Z^d$. In turn, this implies that $Z^s\indep D(\cdot)\mid Z^d$. 
 
H\"{u}nermund and Bareinboim's contribution to this Special Issue provides a detailed exposition of Pearl's approach and a perspective on applying it to current problems in econometrics.

\subsection{Adding more realism through nonparametrics} 
%--------------------------------------------------------------------------------------------------------------------------------------------------
Functional form assumptions play a critical role in Wright's analysis.  The demand and supply elasticities are not identified in fully nonlinear and nonseparable models.  For example, suppose 
\begin{equation}
\label{eqn:nonparDp}
D(p) = \varphi (p, Z^d,W,\boldsymbol{\epsilon}^d),
\end{equation}
where $\boldsymbol{\epsilon}^d$ is a multivariate structural disturbance of unknown dimension and $\varphi$ is a nonparametric structural function. A fundamental result of Pearl shows that such models are not identified nonparametrically.  This means that linear functional forms and additive errors play a strong role in Wright's analysis.  There has been some work in econometrics on relaxing these assumptions.  Newey and Powell (2003) provide identification results for the case 
$$
\varphi(p, Z^d,W,\epsilon^d)= \varpi (p,Z^d,W) + \epsilon^d,
$$ 
where $\boldsymbol{\epsilon}^d=\epsilon^d$ is a scalar additive disturbance, and $\varpi$ is nonparametric.\footnote{Newey, W. K. and J. L. Powell  (2003). Instrumental variable estimation of nonparametric models. {\em Econometrica 71}(5), 1565--78.}  If the distribution of $(P,Z^d,W)$ is complete with respect to $(Z^s,Z^d,W)$; that is, 
$$
\Ep[l(P,Z^d,W) |Z^s,Z^d,W] = 0 \implies l(P,Z^d,W) = 0\text{ almost surely}
$$
for each integrable $l$; then $\varpi$ (and thus $\varphi$) is identified nonparametrically. In this framework, using parametric functional forms for $\varpi$ becomes more like a series approximation to the true $\varpi$. Inference is further refined by imposing shape restrictions such as monotonicity.\footnote{See Chetverikov, D. and D. Wilhelm (2017). Nonparametric instrumental variable estimation under monotonicity. {\em Econometrica 85}(4), 1303--20.}

Further in this vein, Chen {\em et al.} (2014) provide identification results for the nonadditive case \eqref{eqn:nonparDp} with scalar $\boldsymbol{\epsilon}^d=\epsilon^d$,
$$
D(p)=\varphi (p,Z^d,W, \epsilon^d),
$$ 
requiring that $\epsilon^d$ is uniformly distributed on the unit interval and that $\varphi$ is strictly increasing in its last argument.\footnote{Chen, X., V. Chernozhukov, S. Lee and W. K. Newey (2014). Local identification of nonparametric and semiparametric models. {\em Econometrica 82}(2), 785--809.}

If we abandon the assumption that $\boldsymbol{\epsilon}^d$ is scalar, we need to perform bounds analysis; see Chesher and Rosen (2017) and their contribution to this Special Issue.\footnote{Chesher, A. and A. M. Rosen (2017). Generalized instrumental variable models. {\em Econometrica 85}(3), 959--89.}   We can also focus on identifying certain weighted average structural derivatives, similarly to the local average treatment effect of Angrist and Imbens (1994).\footnote{Imbens, G. W. and J. D. Angrist (1994). Identification and estimation of local average treatment eﬀects. {\em Econometrica 62}(2), 467--75.}  This approach faces the caveat that supply shifters generate the weights, so they may not have intrinsic value for purposes of counterfactual analysis.

 \subsection{Giving up realism for formal calibration}   
 %--------------------------------------------------------------------------------------------------------------------------------------------------
 An alternative approach is to give up on modelling the true structural data-generating processes and pursue calibration. In this approach,  we can interpret parametric models such as Wright's as convenient formal calibrations that can be used for coherent policy analysis and may provide sensible ballpark answers.  This calibration approach recognizes that the equilibrium model is an abstraction that provides a convenient framework for analysis; focusing on one good or several goods at a time is another convenient abstraction; a linear in logs demand model is yet another; and so on.  This abstraction is an approximation that may perform well or poorly  in answering policy questions. While the convenience of this approach is clear, there does not appear to be any systematic way to evaluate the policy performance of calibrated models. See Berry and Haile (2021) for a discussion of this tradeoff between convenience and reality  and Sims (1980) for a macroeconometric angle on these issues.\footnote{Berry, S. T. and P. A. Haile (2021). Foundations of demand estimation. In K. Ho, A. Hortacsu and A. Lizzeri (Eds.), {\em Handbook of Industrial Organization}, vol. 4, 1--62. Amsterdam: North-Holland; and Sims, C. A. (1980). Macroeconomics and reality. {\em Econometrica 48}(1), 1--48.}

\subsection{Instruments and proxy controls} 
%--------------------------------------------------------------------------------------------------------------------------------------------------
Among the most interesting recent developments in IV methods is the work on proxy controls.  Roughly speaking, the causal model is identified by a regression of some outcome $Y$ on the policy variable $T$ and latent control $A$ (for example, in returns to education, $Y$ is earnings, $T$ is years of schooling, and $A$ is latent ability).  Multiple measurements of $A$ are available, say $V$ and $W$.  Then the average causal effect of $T$ on $Y$ is identified by an instrumental regression of $Y$ on $T$ and $W$, instrumenting $W$ with $V$, provided that the distribution of $A$ is complete with respect to $V$.  This approach goes back to Zvi Griliches's and Gary Chamberlain's work on linear models.\footnote{See  Griliches, Z. (1977). Estimating the returns to schooling: some econometric problems. {\em Econometrica 45}(1), 1--22; and Chamberlain, G. (1977). Education, income, and ability revisited. {\em Journal of Econometrics 5}, 241--57.}  For nonparametric models, this approach is developed by Tchetgen Tchetgen, Deaner, and others.\footnote{See {\em e.g.} Miao, W., Z. Geng and E. J. Tchetgen Tchetgen (2018). Identifying causal effects with proxy variables of an unmeasured confounder. {\em Biometrika 105}(4), 987--93; Deaner, B. (2023). Proxy controls and panel data. arXiv:1810.00283 {\em [econ.EM]}.}  Kallus, Mao, and Uehara (2022) drop completeness requirements for some policy functionals.\footnote{Kallus, N., X. Mao and M. Uehara (2022). Causal inference under unmeasured confounding with negative controls: a minimax learning approach. arXiv:2103.14029 {\em [stat.ML]}.}

 \section{Outline of the special issue}
 \label{s:outline}
 %===================================================================================
 
The first article following this editorial is our reproduction of Wright's (1928) Appendix B. We have newly typeset the text, including its graphs and table. Along the way, we have corrected four obvious and potentially confusing errors in its mathematical expressions. We have also made various minor typographical changes to improve readability. Our editorial statement on the article's title page accounts for all differences with the original. 

The remaining three articles provide state-of-the-art perspectives on causal inference using DAGs and IVs in econometrics. 

Paul H\"{u}nermund and Elias Bareinboim depart from the research in computer science and statistics that emanated from Judea Pearl's work on nonparametric structural causal models represented by DAGs.\footnote{See Pearl (2009b) in Footnote \ref{fn:cyclical}.} Despite its roots in early econometrics and application in other fields, this work has so far made few inroads in empirical economics. Paul and Elias first provide a brief introduction to the algorithmic causal inference procedures this literature offers to researchers willing and able to code their causal understanding of an applied problem in a DAG. Then, they argue that modern econometrics should embrace these procedures, by presenting their possible applications to three problems in empirical economics: confounding, sample selection, and transporting causal knowledge across environments.  

Next, Kirill Borusyak, Peter Hull, and Xavier Jaravel review design-based identification with formula instruments. Formula instruments combine exogenous shocks with other predetermined variables using a known formula. A common example from empirical economics are shift-share instruments, which embody observed differences in individual exposure to aggregate shocks. 

Finally, Andrew Chesher and Adam Rosen discuss the extension of IV methods beyond Wright's (1928) linear setup, with a focus on models that, unlike the linear one, cannot be inverted to express unobserved heterogeneity as a (single-valued) function of observed variables. This situation, in which the inverted model maps data into sets of unobservables commonly arises in empirical economics. Examples include discrete choice, censored outcome, and random coefficient models. 

 {\em The Econometrics Journal} ensures that all articles it publishes, including those in special issues, are peer reviewed. We would like to warmly thank the referees who provided us with fast and high-quality feedback.\\

 \begin{flushright}    
 Jaap H. Abbring\\
 Managing editor\\
   {\it Tilburg University}\\
   {\it Tilburg, The Netherlands}\\
   Email: {\tt jaap@abbring.org}\\
   \ \\
Victor Chernozhukov\\
Co-editor (until April 2024)\\
   {\it Massachussetts Institute of Technology}\\
   {\it Cambridge, MA, USA}\\
   Email: {\tt vchern@mit.edu}\\
   \ \\
 Iv\'{a}n Fern\'{a}ndez-Val\\
 Associate editor (until Dec. 2023)\\
   {\it Boston University}\\
   {\it Boston, MA, USA}\\
   Email: {\tt ivanf@bu.edu}
   \end{flushright} 
 
\end{document}